\documentclass[runningheads,hidelinks]{llncs}

\usepackage{float}
\usepackage{graphicx}
\usepackage{orcidlink}
\usepackage{todonotes}
\usepackage{svg}
\usepackage{booktabs}
\usepackage{multicol}

\usepackage[acronym]{glossaries}

\newacronym{DB}{DB}{Data Block}
\newacronym{FB}{FB}{Function Block}
\newacronym{GUI}{GUI}{Graphical User Interface}
\newacronym{IDE}{IDE}{Integrated Development Environment}
\newacronym{IoT}{IoT}{Internet of Things}
\newacronym{IIoT}{IIoT}{Industrial IoT}
\newacronym{IT}{IT}{Information Technology}
\newacronym{KG}{KG}{Knowledge Graph}
\newacronym{LOC}{LOC}{Lines of Code}
\newacronym{ML}{ML}{Machine Learning}
\newacronym{NN}{NN}{Neural Network}
\newacronym{OT}{OT}{Operational Technology}
\newacronym{PLC}{PLC}{Programmable Logic Controller}
\newacronym{TinyML}{TinyML}{Tiny Machine Learning}
\newacronym{TIA}{TIA}{Totally Integrated Automation}
\newacronym{TM NPU}{TM NPU}{Technology Module Neural Processing Unit}
\newacronym{UDT}{UDT}{User-Defined Type}
\newacronym{W3C}{W3C}{World Wide Web Consortium}
\newacronym{TD}{TD}{Thing Description}
\newacronym{SOSA}{SOSA}{Sensor, Observation, Sample, and Actuator Ontology}
\newacronym{S3N}{S3N}{Semantic Smart Sensor Network Ontology}
\newacronym{ONNX}{ONNX}{Open Neural Network Exchange}
\newacronym{SSN}{SSN}{Semantic Sensor Network Ontology}
\newacronym{WoT}{WoT}{Web of Things}
\newacronym{OPC UA}{OPC UA}{OPC Unified Architecture}
\newacronym{SeLoC-ML}{SeLoC-ML}{\textbf{Se}mantic \textbf{Lo}w-\textbf{C}ode Engineering for
\textbf{ML} Applications}

\begin{document}

\title{SeLoC-ML: Semantic Low-Code Engineering for Machine Learning Applications in Industrial IoT}

\author{Haoyu Ren\inst{1, 3}\orcidlink{0000-0002-0241-6507} \and
Kirill Dorofeev\inst{1}\orcidlink{0000-0002-5010-281X} \and
Darko Anicic\inst{1}\orcidlink{0000-0002-0583-4376} \and
Youssef Hammad\inst{1, 3}\orcidlink{0000-0001-6622-1971} \and
Roland Eckl\inst{2} \and
Thomas A. Runkler\inst{1, 3}\orcidlink{0000-0002-5465-198X}}
\titlerunning{Semantic Low-Code Engineering for ML Applications in Industrial IoT}
\authorrunning{H. Ren et al.}
% If there are more than two authors, 'et al.' iSs used.
%
\institute{Siemens AG, Otto-Hahn-Ring 6, 81739 Munich, Germany\\
% \institute{Siemens AG, Germany\\
\email{\{haoyu.ren, kirill.dorofeev, darko.anicic, youssef.hammad, thomas.runkler\}@siemens.com}\\
\and
Siemens AG, Siemenspromenade 1, 91058 Erlangen, Germany
\email{eckl.roland@siemens.com}\\
\and
Technical University of Munich, Arcisstr. 21, 80333 Munich, Germany
}
\maketitle              % typeset the header of the contribution
%
%
%
%
% \input{abstract.tex}

% \vspace{-.5cm}
\begin{abstract}
\gls{IoT} is transforming the industry by bridging the gap between \gls{IT} and \gls{OT}.
Machines are being integrated with connected sensors and managed by intelligent analytics applications, accelerating digital transformation and business operations. 
Bringing \gls{ML} to industrial devices is an advancement aiming to promote the convergence of \gls{IT} and \gls{OT}.
However, developing an \gls{ML} application in \gls{IIoT} presents various challenges, including hardware heterogeneity, non-standardized representations of \gls{ML} models, device and \gls{ML} model compatibility issues, and slow application development.
Successful deployment in this area requires a deep understanding of hardware, algorithms, software tools, and applications. 
Therefore, this paper presents a framework called \gls{SeLoC-ML}, built on a low-code platform to support the rapid development of \gls{ML} applications in \gls{IIoT} by leveraging Semantic Web technologies. 
\gls{SeLoC-ML} enables non-experts to easily model, discover, reuse, and matchmake \gls{ML} models and devices at scale. 
The project code can be automatically generated for deployment on hardware based on the matching results. 
Developers can benefit from semantic application templates, called \textit{recipes}, to fast prototype end-user applications. 
The evaluations confirm an engineering effort reduction by a factor of at least three compared to traditional approaches on an industrial \gls{ML} classification case study, showing the efficiency and usefulness of \gls{SeLoC-ML}. 
We share the code and welcome any contributions\footnote{\raggedright \url{https://github.com/Haoyu-R/SeLoC-ML}}.
\end{abstract}
%   \vspace{-1cm}
\glsresetall

\section{Introduction}

One of the biggest challenges in industrial digitization is to bridge the gap between \gls{OT} and \gls{IT}. \gls{OT} is centered on a physical world composed of machines, manufacturing equipment, and other hardware, where a massive amount of data is generated. However, \gls{IT} is focused on the contemporary digital world, using data centers, servers, and smart applications to consume the data. These two domains have traditionally functioned in isolation~\cite{Garimella2018}. The rise of Industry 4.0, along with increasing connectivity between humans, machines, and sensors, is driving the convergence of \gls{IT} and \gls{OT}, shifting data-supported decision-making from the individual to the system level and enhancing factory efficiency. 
However, \gls{IT}/\gls{OT} convergence is difficult to achieve. One example is the deployment of \gls{ML} on industrial devices, where \gls{ML} presents the \gls{IT} world and industrial devices present the \gls{OT} world.

\gls{ML} is one of the fast-growing technical advancements. Applying \gls{ML}, specifically \gls{NN}, in the industry by leveraging sensor and system data can provide reliable insights into the factory and accelerate smart manufacturing. Standard \gls{ML} applications transfer massive field data to the cloud and centrally process the data against \gls{NN} models. Concerns have been raised because this data transfer causes numerous issues, such as high energy consumption and latency, privacy leaks, bandwidth congestion.

With the \gls{IoT} growth, factories will be equipped with increasingly powerful, connected, and intelligent devices. This plays a key role in the continuing industrial evolution.
% It is estimated that the global shipment of microcontrollers adds up to 30 billion per year\footnote{\raggedright \url{https://www.statista.com/statistics/935382/worldwide-microcontroller-unit-shipments}}. 
% The proliferation of \gls{IoT} devices will play a key role in the continuing industrial evolution. 
% \gls{TinyML} is an emerging field that democratizes \gls{ML} on \gls{IoT} devices, which 
Offloading \gls{ML} intelligence from the cloud to the \gls{IIoT} devices enables performing \gls{ML} tasks near data sources and reducing reliance on data transfer, which addresses the latency and security concerns.
However, applying on-device \gls{ML} in the industry where mass deployment happens is still challenging.

\gls{IIoT} devices are specialized to fulfill different tasks. 
They come in all shapes and sizes, differ in terms of onboard sensors, available memory and storage capacities, and have various runtime platforms. 
% They have various onboard sensors to monitor the environment. 
% They are designed with different computational resources to optimize energy consumption. 
In the context of on-device \gls{ML}, they rely on \gls{NN} models to interpret sensor data, make predictions about their environments, and take intelligent actions locally. 
\gls{NN} models are developed with various structures, e.g., different combinations of layers and individualized pre- and postprocessing blocks. Additionally, most trained \gls{NN} models are distributed as binary files without a clear and standardized description of their usages. The diversification and proliferation of hardware (\gls{IIoT} devices) and software (\gls{NN} models) widen the gap between each other. 

Many compatibility issues must be carefully investigated to run \gls{ML} properly on the devices, such as sensor input format, memory constraints, and sensor availability. Specifically, we want to answer two sets of questions: 

\begin{enumerate}
    \item How do we achieve the co-management of \gls{IIoT} devices and \gls{NN} models?
    \begin{enumerate}
        \item How do we determine which devices may execute a specific \gls{NN}?
        \item Given a device, how do we determine which trained \gls{NN} model is compatible with it? Does the model meet requirements for accuracy, memory, and latency?
    \end{enumerate}
    \item How do we accelerate the engineering and deployment of \gls{ML} applications in \gls{IIoT}? How might cross-domain collaborations be facilitated and the solution be made accessible to all?
    % \begin{enumerate}
    %     \item How might cross-domain collaborations be facilitated and the solution be made accessible to all?
    %     \item How should the solution be scaled?
    % \end{enumerate}
\end{enumerate}

We present a framework called \gls{SeLoC-ML}, a system for managing and deploying \gls{ML} on devices in \gls{IIoT} at scale. 
Here, we propose to use formalized semantic models to describe heterogeneous \gls{IIoT} devices and \gls{NN} models, respectively. 
With ontology schemas, we can model the knowledge about devices and \gls{NN} models semantically in a unified language and centrally store it in a \gls{KG}, making the knowledge searchable like Web resources. 
As a result, many features are enabled, such as vendor-agnostic knowledge discovery and matchmaking model requirements with device capabilities.

Another aspect of our work is deploying \gls{NN} models on the devices and integrating \gls{ML} applications into end-user \gls{IIoT} applications. 
More crucially, we aim to make the approach accessible and understandable to non-experts.
% many intelligent solutions of the future will be invented by innovators with no formal training in semantics or \gls{TinyML}. 
We propose integrating \gls{SeLoC-ML} into a low-code platform, allowing developers without necessary expertise to use semantic services and deploy \gls{ML} models declaratively, advancing \gls{IT}/\gls{OT} convergence.
In the background, user inputs are parsed, and corresponding SPARQL queries\footnote{\raggedright \url{https://www.w3.org/TR/rdf-sparql-query}} are formulated to retrieve the information from the central graph database automatically.
Additionally, different deployment options will become available depending on matching results, allowing developers to generate projects and deploy \gls{ML} on the devices with minimal effort. 
Last but not least, we leverage semantic application templates, so-called \textit{recipes}~\cite{Thuluva2017}, to assist developers in integrating ML applications into greater pipelines and creating end-user applications rapidly.
We support application development by matching the data types used in \textit{recipes} with the data points provided by the devices, which are defined by common semantic models.

As an example of this approach, we present the solution on a Siemens \gls{PLC} SIMATIC S7-1500\footnote{\raggedright \url{https://new.siemens.com/global/en/products/automation/systems/industrial/plc/simatic-s7-1500.html}} using the Siemens low-code platform Mendix\footnote{\raggedright \url{https://www.mendix.com}}~\cite{Litman2018}. We demonstrate in Mendix how to search and matchmake an \gls{NN} model with a SIMATIC S7-1500 \gls{TM NPU}\footnote{\raggedright \url{https://new.siemens.com/global/en/products/automation/systems/industrial/plc/simatic-s7-1500/simatic-s7-1500-tm-npu.html}} connected with an Intel RealSense camera\footnote{\raggedright \url{https://www.intel.com/content/www/us/en/architecture-and-technology/realsense-overview.html}}. The goal is to find an trained \gls{NN} model compatible with the \gls{TM NPU} for classifying different types of objects on a conveyor belt. Following a successful match, an engineering project for \gls{TIA} Portal\footnote{\raggedright \url{https://new.siemens.com/global/en/products/automation/industry-software/automation-software/tia-portal.html}} - a Siemens \gls{IDE} for industrial automation - can be automatically created and ready for deployment. Later, we present how to use a semantic \textit{recipe} to orchestrate the \gls{ML} application and easily build an end-user application in Mendix to monitor the classification results. The evaluation results show that \gls{SeLoC-ML} can reduce engineering effort by a factor of at least three compared to conventional approaches. 

The remainder of this paper first presents related work on \gls{ML} in \gls{IoT}, Semantic Web technologies, \gls{ML} management, and low-code programming in Section~\ref{section:related_work}.
Section~\ref{section:approach} describes \gls{SeLoC-ML} from the semantic system to low-code platform integration.
Section~\ref{section:in-use-example} demonstrates \gls{SeLoC-ML} on an industrial classification problem using Siemens products as an example.
In Section~\ref{section:evaluation}, we evaluate the approach, compare it with the traditional workflow and provide the benefits of \gls{SeLoC-ML}. Finally, Section~\ref{section:conclusion} concludes the paper and discusses future work.

\section{Related Work} 
\label{section:related_work}

\subsubsection{Advancements of on-device \gls{ML}}

On-device \gls{ML} is more than just an algorithm. 
It is about the proliferation of hardware, progress on algorithms, emerging ecosystem, and transformative applications. 
Ultra-low-power devices have been designed for always-on applications~\cite{Giordano2021}~\cite{Jiao2021}. 
Various algorithms have been proposed to fully exploit \gls{ML} models on the devices without compromising performance~\cite{Song2022}~\cite{Ren2021}.
Collaborative ecosystems can further squeeze the potential from the synergism of hardware and software~\cite{Duan2022}~\cite{Rashid2022}. 
Last but not least, many applications have been brought from a proof-of-concept to products~\cite{Gomez2022}~\cite{BejaranoCarbo2022}.

\subsubsection{Semantic Web Technologies}

Semantic Web technology provides means for building, storing, and handling diverse data sources of different structures, making it an ideal candidate for information modeling and integration in \gls{IoT}. 
Evidence has shown the benefits of semantics in industrial domains~\cite{Rojas2021}~\cite{Mehdi2017}. 
In the context of \gls{IoT}, ontologies like \gls{SOSA}~\cite{Janowicz2019}, \gls{SSN}~\cite{Compton2012}, and \gls{S3N}~\cite{Sagar2018} are few prominent semantic models for describing intelligent IoT devices, their properties, and interactions. 
The \gls{TD}~\cite{Charpenay2016} ontology developed by the \gls{W3C} \gls{WoT}\footnote{\raggedright \url{https://www.w3.org/WoT}} working group specifies the metadata and interfaces of \gls{IoT} devices.
\emph{iotschema.org}\footnote{\raggedright \url{http://iotschema.org/docs/full.html}} is yet another semantic model in the \gls{IoT} domain, which is used to enrich the data among connected things. 
This study is interested in combining these semantic schemas with our proposed \gls{NN} model ontology~\cite{Ren2022} for modeling the heterogeneous knowledge about devices and \gls{NN} models in \gls{IIoT}.

\subsubsection{Management of on-device \gls{ML}}
There are hundreds of billions of IoT devices today, and new \gls{ML} models are developed and distributed daily. 
To manage these resources at scale, it is necessary to increase the interoperability and transparency of the ecosystem. 
\gls{ONNX}\footnote{\raggedright \url{https://onnx.ai}} aims to provide a shared exchange format that allows developers to use \gls{ML} models across different deep learning frameworks. However, it fails to provide descriptions of models in a formal way. To overcome this limitation, TensorFlow Lite Metadata\footnote{\raggedright \url{https://www.tensorflow.org/lite/convert/metadata}} and Model Card~\cite{Mitchell2019} are introduced to formally document \gls{ML} models. Few databases~\cite{Vartak2016}~\cite{Nguyen2020} are introduced for tracking \gls{ML} models. Nevertheless, they do not scale well since many manual works are required, and their information models do not express the relationships between \gls{ML} models and hardware. 
\gls{ML} models need to be studied together with the specifications of hardware to achieve joint management. 

\subsubsection{Low-Code Engineering}

Despite remarkable \gls{IT}/\gls{OT} integration achievements, the current state of developing complex \gls{IIoT} applications is still far from satisfactory~\cite{udoh2018developing}.
The concept of low-code engineering and corresponding platforms, such as Mendix~\cite{Litman2018}, are introduced to support fast application development without a prerequisite of having enhanced coding skills~\cite{ihirwe2020low}.
Low-code concepts find their applications in the manufacturing domain~\cite{Sanchis2020}~\cite{waszkowski2019low}, allowing to quickly build industrial applications based on the services provided by the machines on the shop floor.
To match the business requirements with the existing functionalities of the machines and compose them meaningfully, we use the notion of \textit{recipes}~\cite{Thuluva2017} as an easy way to model such compositions.
\textit{Recipes} can be seen as application templates~\cite{Thuluva2020} developed to solve a class of problems and can be later easily configured for a specific use case.

\section{Approach}
\label{section:approach}

We present \gls{SeLoC-ML} considering the interoperability and deployment obstacles in \gls{ML} applications in \gls{IIoT}. 
This section starts by introducing the framework setup. 
The proposed architecture relies on a semantic system designed to cover but not be limited to the use case addressed in this study.
We illustrate the semantic system, from the ontology to the semantic services. 
Next, a simplified \gls{KG} and two SPARQL queries are presented to exemplify the system's advantages. 
We then propose integrating \gls{SeLoC-ML} into the Siemens Mendix low-code platform allowing developers to easily identify and matchmake components, deploy \gls{NN} models to the devices upon matching, and quickly prototype \gls{IIoT} user applications.

\subsection{Framework Architecture}

\begin{figure*}[t]
      \centering
      \includegraphics[width=0.77\linewidth]{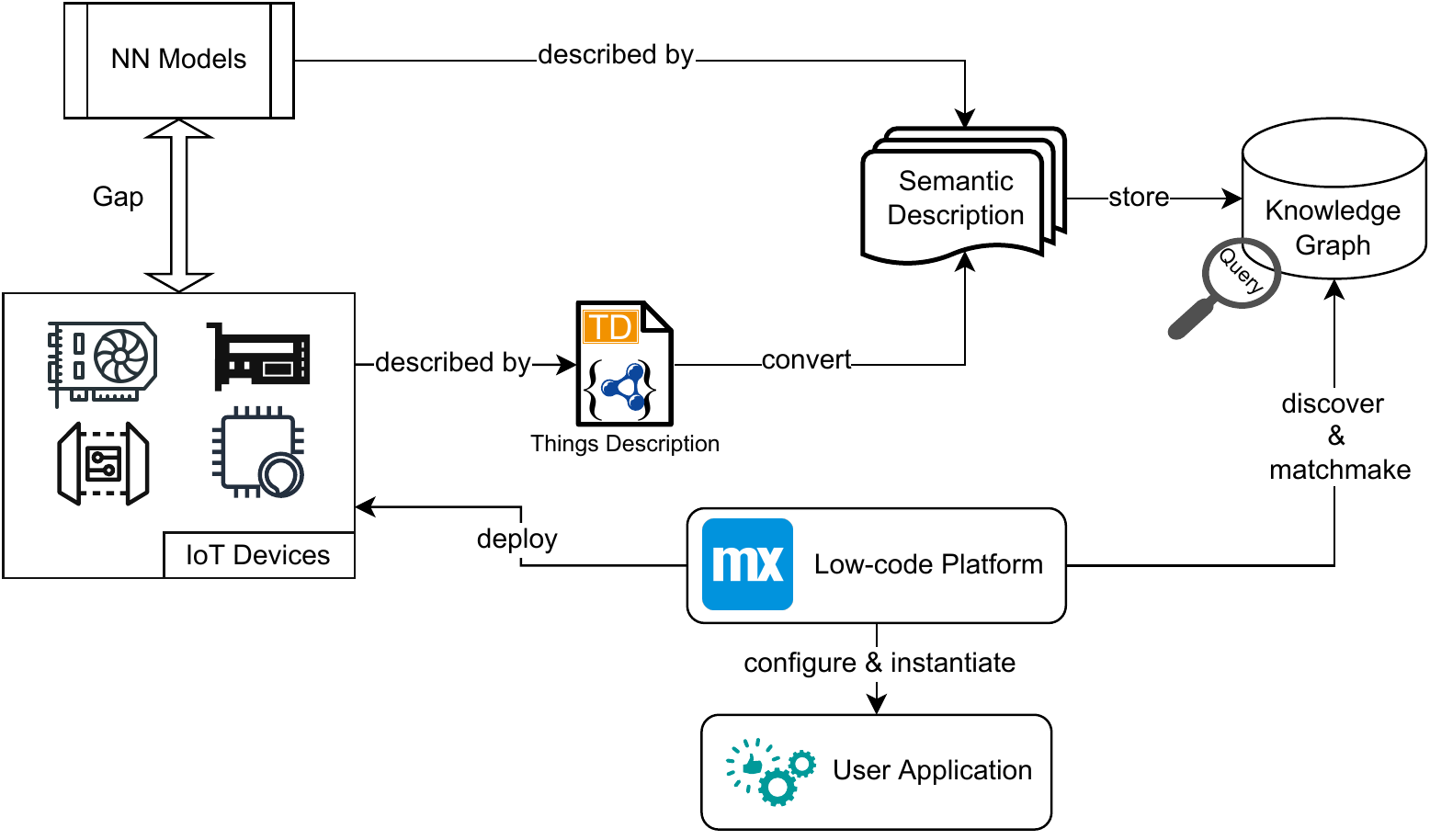}
      \caption{Framework architecture of \gls{SeLoC-ML}.}
      \label{Framework_Pic}
      \vspace{-0.4cm}
\end{figure*}

Figure~\ref{Framework_Pic} presents the \gls{SeLoC-ML} framework. 
The figure on the left illustrates that developers are faced with a gap between software (\gls{NN} models) and hardware (\gls{IIoT} devices).
We first propose to utilize two semantic models to describe \gls{IIoT} devices and \gls{NN} models, respectively. 
Here, any formalized semantic models can be applied, but for demonstrating the technology in the industrial environment, we choose the standardized \gls{W3C} \gls{TD}~\cite{Charpenay2016} to describe devices. 
Aligning with the \gls{TD}, we design a semantic ontology~\cite{Ren2022} with a conversion tool for describing \gls{NN} models in terms of their metadata, structures, and hardware requirements.
Thus, knowledge about heterogeneous \gls{IIoT} devices and \gls{NN} models can be translated into unified semantic descriptions against their ontologies and be hosted together in a \gls{KG}, as shown on the right side of the figure. 
The bottom of the figure shows that even non-experts can easily scrape the \gls{KG} with Mendix.
Mendix will automatically formulate queries based on user inputs and retrieve desired answers from the graph.
Upon matchmaking, different deployment options are made available. 
A ready-to-be-deployed engineering project can be generated based on user configurations and the retrieved semantic information. 
This is known as \gls{ML}-as-a-service~\cite{Doyu2021}.
Finally, developers can leverage semantic application templates \textit{recipes} to accelerate user application development.

\subsection{Semantic System}
\subsubsection{Ontology}

\begin{figure*}[t]
      \centering
      \includegraphics[width=0.90\linewidth]{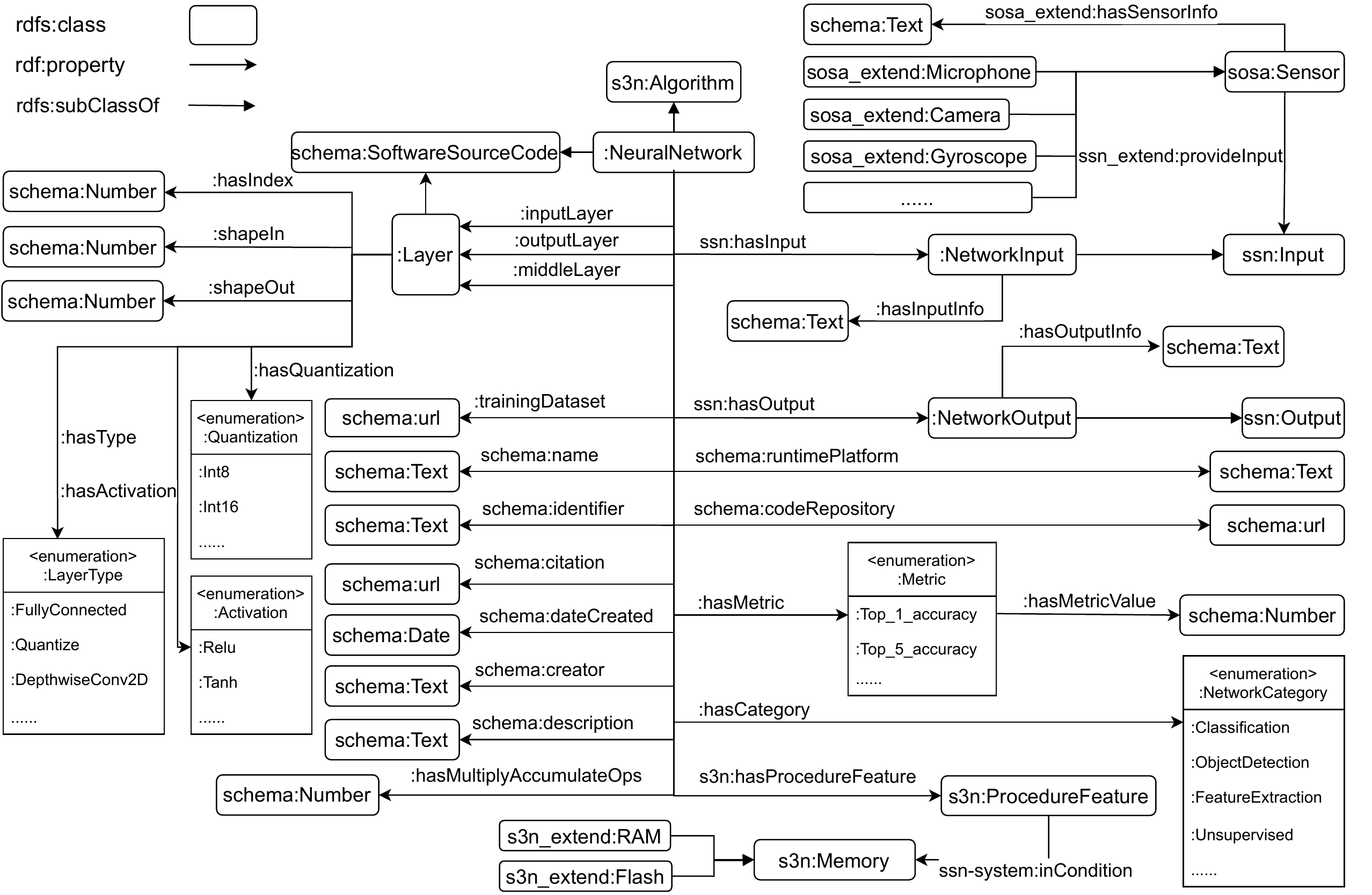}
      \caption{Ontology of \gls{NN} model in \gls{IoT}.}
      \label{Ontology_Pic}
      \vspace{-0.4cm}
\end{figure*}

We presented an ontology\footnote{\raggedright \url{https://tinyml-schema-collab.github.io}}~\cite{Ren2021} to describe \gls{NN} models in the context of \gls{IoT}, as shown in Figure~\ref{Ontology_Pic}. 
By reusing existing schemas, such as \gls{S3N} and \gls{SOSA}, we aligned the ontology with other Web standards and avoided reinventing the wheel.
For research and demonstration purposes, the ontology has been designed to guarantee its interoperability and compatibility with \gls{TD}, which we applied to describe \gls{IIoT} devices.
The ontology can render three different forms of information about a \gls{NN}: 1)~metadata, such as the date of creation, category, and literal description; 2)~structure, such as the input and output layers; 3)~hardware requirements, such as memory and sensors.

As references, we provide interested readers with examples of semantic descriptions of \gls{IoT} devices and \gls{NN} models in our repository. Additionally, scripts are available, which can generate a semantic representation of a given \gls{NN} model along with some user inputs since not all information can be obtained by parsing the \gls{NN} model, such as dataset and author information.

\subsubsection{Knowledge Graph}

\begin{figure*}[t]
      \centering
      \includegraphics[width=0.95\linewidth]{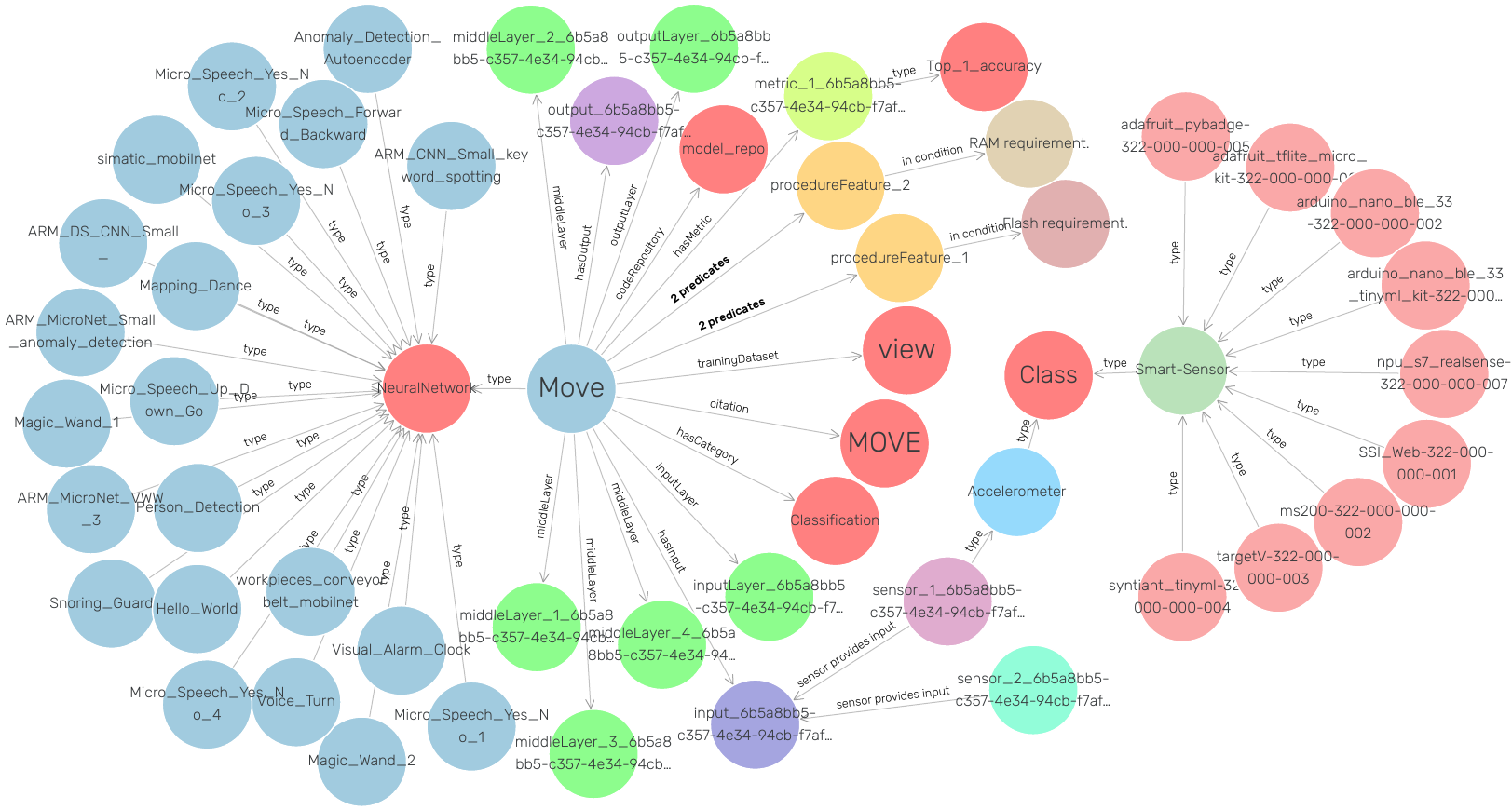}
      \caption{A simplified \gls{KG} containing 22 \gls{ML} models and nine devices.}
      \label{KG_Pic}
      \vspace{-0.4cm}
\end{figure*}

We can design a central \gls{KG} that stores information from \gls{NN} models and \gls{IoT} devices using the semantic schemas introduced above.
For an example, we used GraphDB\footnote{\raggedright \url{https://www.ontotext.com/products/graphdb}} to demonstrate a simplified \gls{KG} composed of nine \gls{IoT} devices and 22 \gls{NN} models. As depicted in Figure~\ref{KG_Pic}, the \gls{NN} model and device nodes are drawn on the left and right sides of the figure, respectively. An \gls{NN} model called \textit{Move} is expanded in the center, displaying its properties. We collect \gls{NN} models trained using TensorFlow, one of the most prominent deep learning frameworks. However, our approach can be easily scaled to cover different devices, \gls{NN} models, and frameworks. As previously mentioned, we provide the code and examples for creating \gls{KG} and interacting with it. 

\subsubsection{Discovery and Matchmaking}

Many specific uses and services can be enabled once the \gls{KG} has been created. We use two simple queries to answer two questions against the \gls{KG} example introduced above for a demonstration. More queries can be found in our repository. The used namespaces and corresponding prefixes are given as follows:

\begin{multicols}{2}[\scriptsize]
\begin{verbatim} 
# Our NN ontology
nnet: <http://tinyml-schema.org/
      networkschema/>
# Schema.org Vocabulary
schema: <https://schema.org> .
# Units of Measure Vocabulary
om: <http://www.ontology-of-units-of
     -measure.org/resource/om-2/> .
# SSN Ontology
ssn: <http://www.w3.org/ns/ssn/> .

# S3N Ontology
s3n: <http://w3id.org/s3n/> .
# Extension of the SOSA ontology
sosa_extend:  <http://tinyml-schema.org/
              sosa_extend#> .
# Extension of the SSN ontology
ssn_extend:  <http://tinyml-schema.org/
             ssn_extend#> .
# Extension of the S3N ontology
s3n_extend:  <http://tinyml-schema.org/
             s3n_extend#> .
\end{verbatim}
\end{multicols}

\begin{enumerate}
    \item We have an \gls{IoT} device on which we want to deploy an \gls{NN} model. The device is equipped with a camera, and it has 144 and 621 Kb of available RAM and Flash, respectively. We want to determine all possible \gls{NN} models that can be executed on this device.
    \item We trained an \gls{NN} model for motion classification using gyroscope and accelerometer data. Given that the minimum RAM and Flash requirements for running this model are 121 and 610 Kb, respectively, we want to know which available devices can run this model.
\end{enumerate}

% \begin{table}[h]
% \renewcommand\tablename{Code}
\begin{multicols}{2}[\scriptsize]
\begin{verbatim}
Query 1:

SELECT ?uuid ?MACs ?RAM ?Flash ?Description
WHERE {
    ?nn a nnet:NeuralNetwork ;
        schema:identifier ?uuid ;
        schema:description ?Description ;
        ssn:hasInput ?input;
        nnet:hasMultiplyAccumulateOps ?MACs ;
        s3n:hasProcedureFeature ?x_1 ;
        s3n:hasProcedureFeature ?x_2 .
    ?x_1 ssn-system:inCondition ?cond_1 .
    ?x_2 ssn-system:inCondition ?cond_2 .
    ?cond_1 a s3n_extend:RAM ;
        schema:minValue ?RAM ;
        schema:unitCode om:kilobyte .
    ?cond_2 a s3n_extend:Flash ;
        schema:minValue ?Flash ;
        schema:unitCode om:kilobyte .
    ?sensor ssn_extend:provideInput ?input;
        a sosa_extend:Camera .
    FILTER (?RAM <= 144)
    FILTER (?Flash <= 621)
}

Result: 
uuid: 2c... ; MACs: 7158144; RAM: 94 Kb; ...
uuid: 49... ; MACs: 7387976; RAM: 116 Kb; ...

Query 2:

SELECT ?Device ?RAM ?Flash
WHERE {
    ?Device a s3n:SmartSensor ;
        ssn:hasSubSystem ?system_1 ;
        ssn:hasSubSystem ?system_2 ;
        ssn:hasSubSystem ?system_3 .
    ?system_1 a sosa_extend:Accelerometer .
    ?system_2 a sosa_extend:Gyroscope .
    ?system_3 a s3n:MicroController ;
        s3n:hasSystemCapability ?x .
    ?x ssn-system:hasSystemProperty ?cond_1 .
    ?x ssn-system:hasSystemProperty ?cond_2 .
    ?cond_1 a s3n_extend:RAM ;
        schema:value ?RAM ;
        schema:unitCode om:kilobyte .
    ?cond_2 a s3n_extend:Flash ;
        schema:value ?Flash ;
        schema:unitCode om:kilobyte .
    FILTER (?RAM >= 121)
    FILTER (?Flash >= 610)
}
ORDER BY ?RAM

Result: 
Device: 002; RAM: 172 Kb; Flash: 628 Kb.
Device: 003; RAM: 187 Kb; Flash: 785 Kb.
\end{verbatim}
\end{multicols}
%     \caption{My Caption}
%     \label{my-label}
% \end{table}

\subsection{Low-Code Platform Integration}

Semantic Web techniques are not easy to learn and use.
Likewise, on-device \gls{ML} is another entirely different field that is challenging to understand. 
To motivate cross-domain collaborations and simplify IT/OT convergence, we encourage integrating \gls{SeLoC-ML} into a low-code platform – Mendix. Mendix allows developers to design, build, deploy, and operate \gls{IoT} applications rapidly.

\subsubsection{Semantic Management of on-device \gls{ML}}

We created a Mendix application with a user-friendly \gls{GUI} connected with a \gls{KG} in the background. The application package is published in our repository. Three main semantic services are provided in the application:

\begin{figure*}[t]
      \centering
      \includegraphics[width=0.86\textwidth]{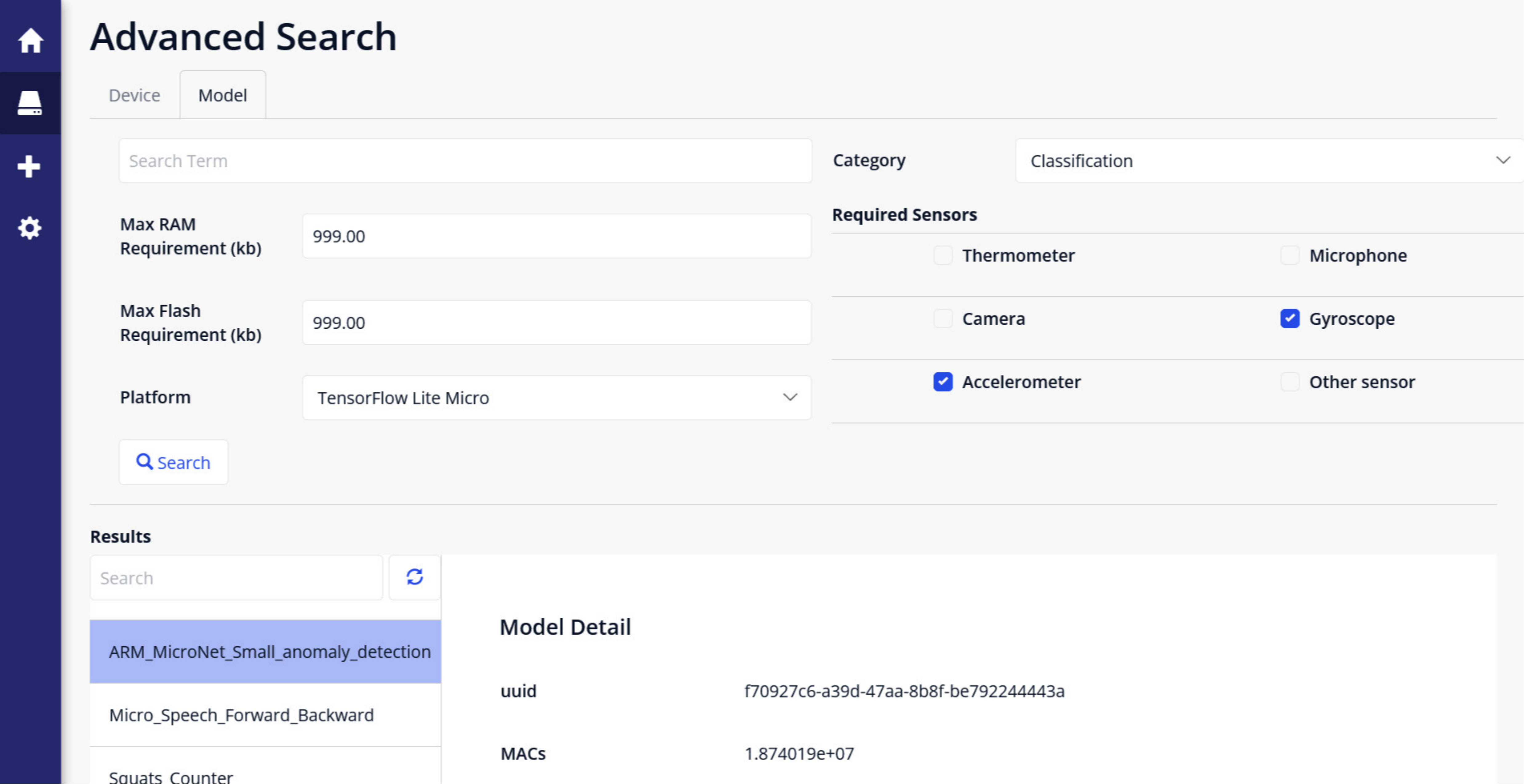}
      \caption{Semantic similarity search.}
      \label{Mendix_Similarity_Pic}
      \vspace{-0.4cm}
\end{figure*}

\begin{enumerate}
    \item \textbf{Discovery}: Developers can browse through all available \gls{NN} models and \gls{IIoT} devices in the graph database and inspect their details.
    \item \textbf{Matchmaking}: Once the developer selects an \gls{IIoT} device/\gls{NN} model in the application, SPARQL queries are automatically formulated to retrieve all compatible \gls{NN} models/devices.
    \item \textbf{Semantic similarity search}: Imagine if hundreds of thousands of \gls{IIoT} devices and \gls{NN} models were hosted in the \gls{KG}, it would be tedious to examine them one after another manually. 
    Semantic similarity search\footnote{\raggedright \url{https://graphdb.ontotext.com/documentation/standard/semantic-similarity-searches.html}} enables users to explore relevant objects in the \gls{KG} by typing a search text, similar to Google Search. 
    In our example, users can search the stored components by filling in the provided form with their requirements, as shown in Figure~\ref{Mendix_Similarity_Pic}. 
\end{enumerate}

\subsubsection{\gls{ML}-as-a-service}

Moreover, depending on the matched devices, different deployment options become available, and corresponding project code can be generated by parsing the retrieved information. 
Of course, specific user configurations will be asked to complete the project creation. We aim to provide a high-level abstraction for deployment, which is as hardware-agnostic as possible.

\subsubsection{Rapid User Applications Development}
Further, we use Mendix to rapidly prototype an \gls{IIoT} application using the concept of \textit{recipes}. 
\textit{Recipes} provide platform-agnostic application templates that can be easily deployed and configured for common automation tasks.
They specify the application logic and the information about input and output required by the tasks~\cite{Thuluva2020}.
Device data is described using standardized semantic models, such as \gls{OPC UA} companion specifications\footnote{\raggedright \url{https://opcfoundation.org/about/opc-technologies/opc-ua/ua-companion-specifications/}}.
Such a semantic model is stored in the knowledge graph and, at the same time, it is used for developing the \textit{recipes} logic in Mendix.
End-user applications can be instantiated by matching the data points provided by system components with the data types required by the \textit{recipe}. 
As a result, developers quickly orchestrate an end-user application, ready to be delivered.

\section{Workflow Comparison and Case Study}
\label{section:in-use-example}

\begin{figure*}[t]
      \centering
      \includegraphics[width=0.65\linewidth]{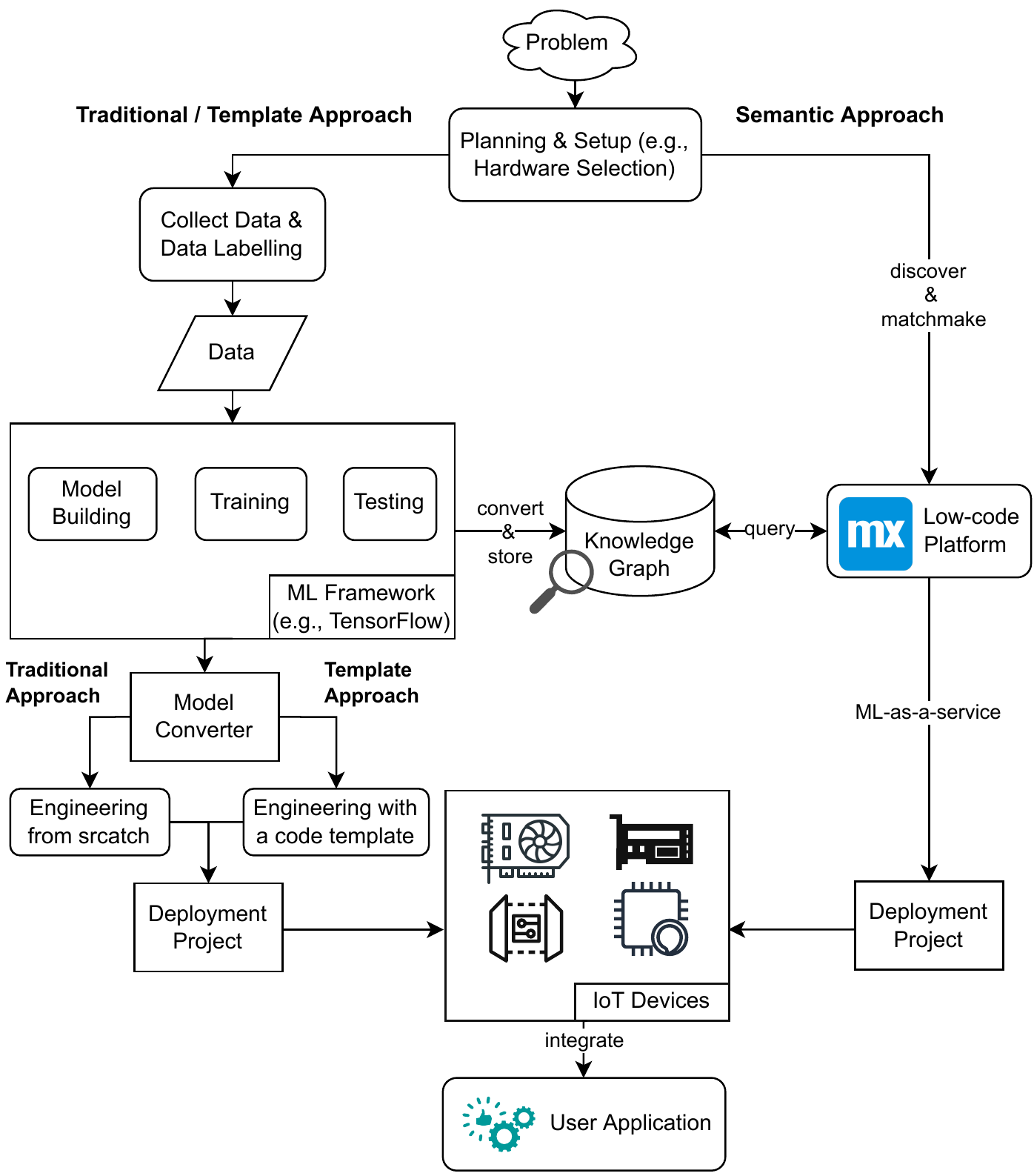}
      \caption{Comparison between the traditional, template, and semantic approach.}
      \label{Workflow_Pic}
      \vspace{-0.4cm}
\end{figure*}

This section first outlines \gls{SeLoC-ML} by comparing it with the State-of-the-Art approaches. Figure~\ref{Workflow_Pic} presents three different workflows for constructing \gls{ML} applications in \gls{IIoT}. 
Traditionally, after the project planning, \gls{ML} developers are engaged to engineer an \gls{ML} model systematically, from data collection and labeling to model building and training. 
% Various measures will be taken to improve the model's performance until it meets the expectations. 
Afterward, embedded engineers take over the work, where the trained model is optimized and converted for the target runtime platform through Model Optimizer.
An embedded project is then engineered with the \gls{ML} model and uploaded to the device. 
Later, software engineers design a user application to integrate the \gls{ML} application and report the results to end-users.
As can be seen, it is difficult to feature an \gls{IIoT} \gls{ML} application that requires cross-domain expertise and a significant amount of engineering work. 

\gls{SeLoC-ML} offers an all-in-one solution based on Mendix low-code platform to alleviate the situation. 
\gls{SeLoC-ML} is generic enough, but for easily quantifying the evaluation and demonstrating its benefits, we illustrate it on an industrial \gls{ML} classification use case where a \gls{NN} model is to be discovered and applied on a Siemens \gls{TM NPU}. 
Of course, the \gls{SeLoC-ML} framework can be quickly scaled and applied to other domains and/or use cases.

\subsection*{In-use: Building an \gls{ML} Application on Siemens SIMATIC}

\begin{figure}
    \centering
	\includegraphics[width=.77\textwidth]{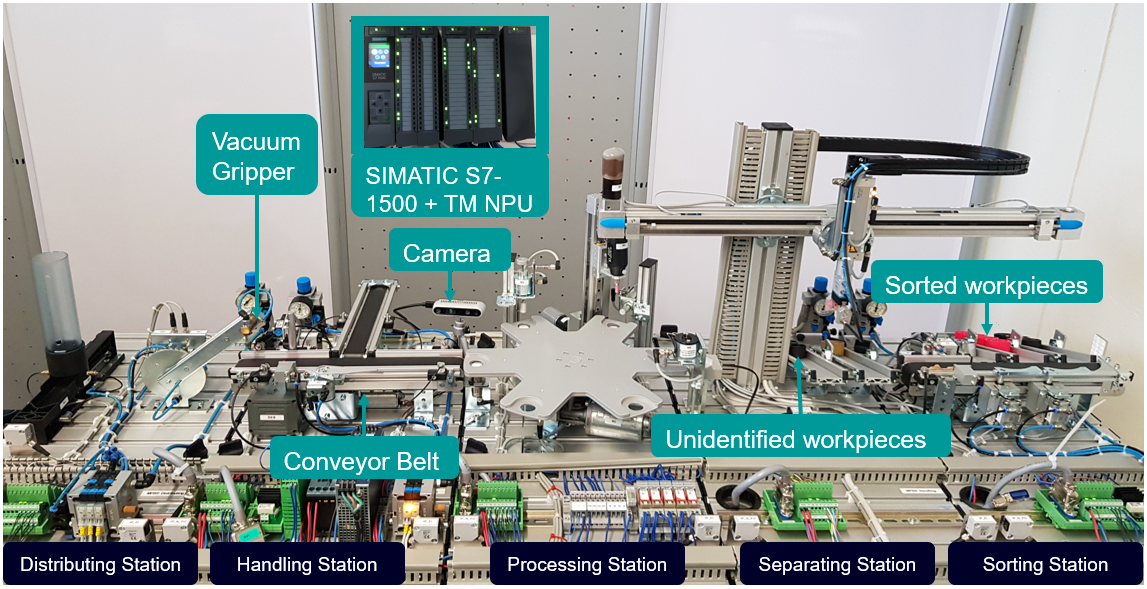}
	\caption{Festo Didactic working station controlled by a SIMATIC S7-1500 \gls{PLC} with a connected \gls{TM NPU}.}
	\label{fig:festo_ws}
      \vspace{-0.4cm}
\end{figure}

The case study is conducted on a Festo Didactic workstation\footnote{\raggedright \url{https://www.festo-didactic.com/int-en/learning-systems/process-automation-control-theory}} controlled by a Siemens SIMATIC S7-1500 \gls{PLC}, as shown in Figure~\ref{fig:festo_ws}. 
In the running example, the vacuum gripper on the left side puts workpieces on the conveyor belt, transporting them to the following process. 
Different workpieces need to be classified for different downstream handlings, and unidentified objects should be sorted out before the next step. 
A \gls{TM NPU} connected with an Intel RealSense camera is installed on the workstation, controlled by the SIMATIC S7-1500 \gls{PLC}.
\gls{TM NPU} enables the execution of ML models directly on Siemens \glspl{PLC}. 
Our goal is to leverage Mendix to discover, configure, and deploy an \gls{NN} model on the \gls{TM NPU} for classifying workpieces on the conveyor belt using images captured by the camera.

\subsubsection{Discovery and Matchmaking}

\begin{figure}
    \centering
	\includegraphics[width=.86\textwidth]{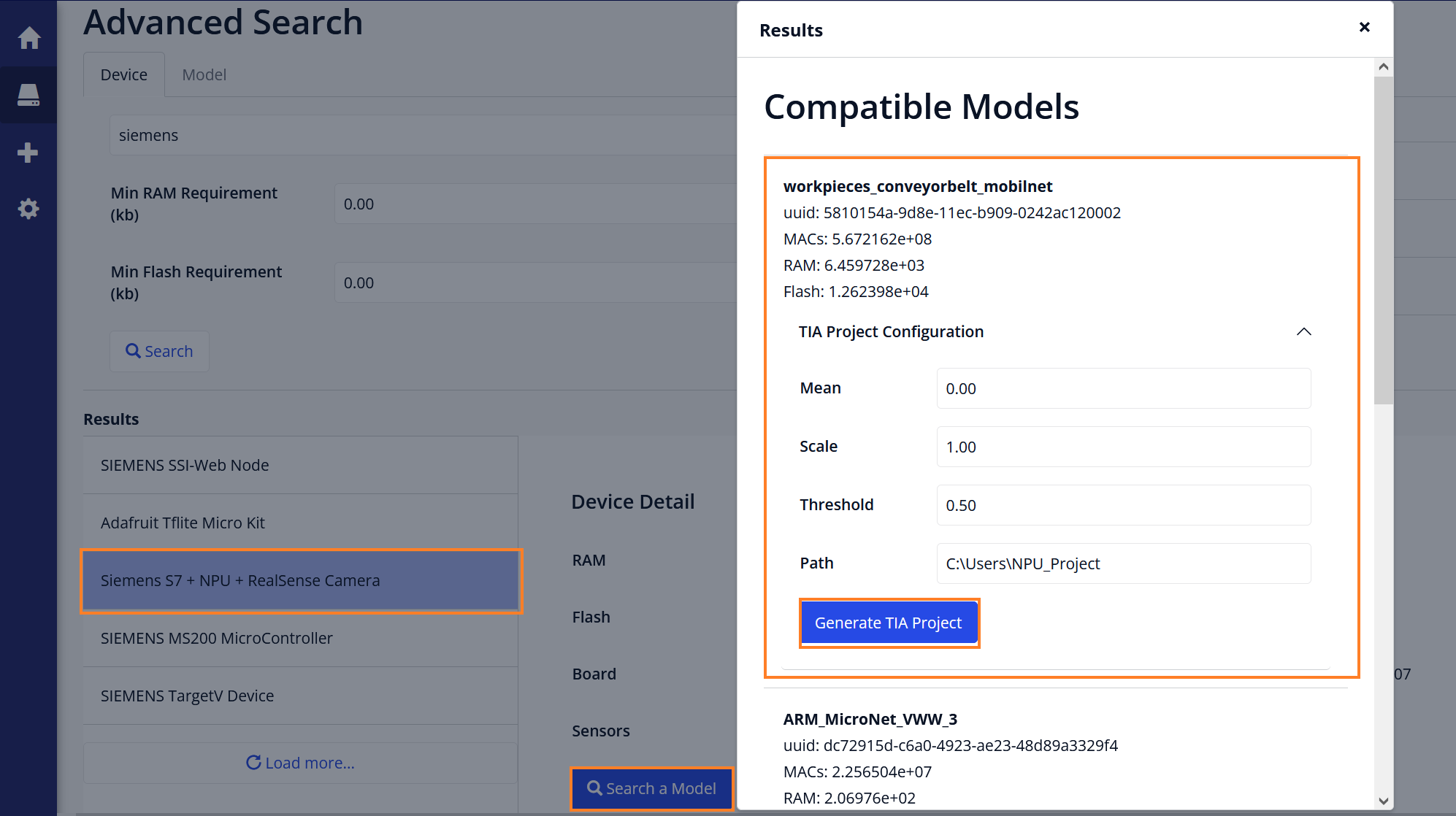}
	\caption{Discovery and matchmaking results.}
	\label{fig:matchmaking}
      \vspace{-0.5cm}
\end{figure}

We explore all the reusable \gls{ML} models in the \gls{KG} that can run on the SIMATIC \gls{TM NPU} without spending much time going through the traditional approach and generating a new model from scratch.  
This can be done in Mendix with a simple click, and all compatible \gls{ML} models will show up in a pop-up window, as depicted in Figure~\ref{fig:matchmaking}. After reviewing the results, we select the model \emph{workpieces\_conveyorbelt\_mobilnet} for our use case.

\subsubsection{Deploying \gls{ML} on Hardware}

After matchmaking, different deployment options become available depending on the selected devices and their runtime platforms. 
In our case, Mendix creates relevant files for the \gls{TM NPU} and \gls{PLC} project. 
As shown in Figure~\ref{fig:matchmaking}, specific configurations still need to be given by users, but most of the information in the project is filled automatically by parsing the retrieved semantic descriptions. 
With all project files loaded to the hardware, the ML application is now ready for execution. 

\subsubsection{Creating a User Application using Recipe}

\begin{figure}
    
    \centering
	\includegraphics[width=.86\textwidth]{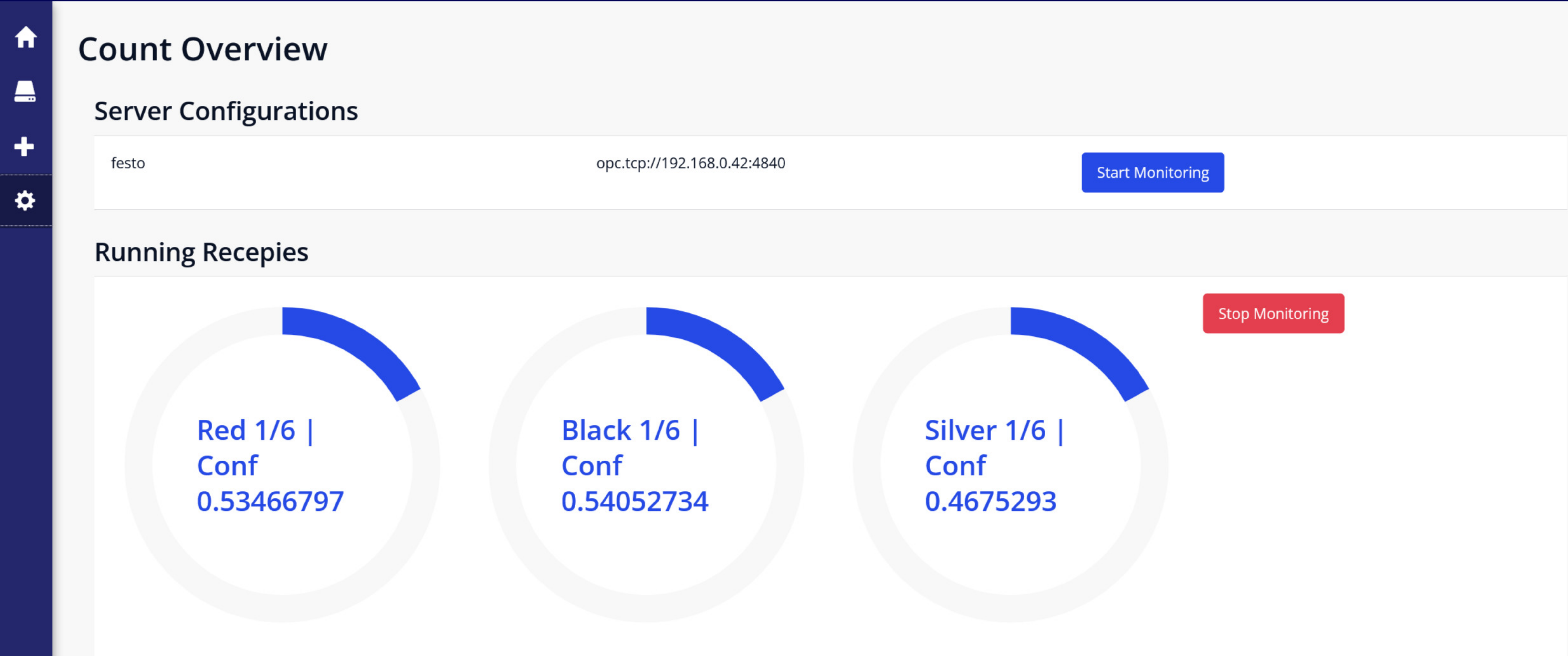}
	\caption{End-user application that monitors the classification results.}
	\label{fig:opcua_recipe}
      \vspace{-0.4cm}
\end{figure}

The classification results from the \gls{TM NPU} are made available via an \gls{OPC UA} server.
We created a \textit{recipe} that provides a template for visualizing object classification results based on their color for our running example. 
To instantiate the \textit{recipe}, Mendix matches the data supplied by the \gls{PLC} with the data types defined in the \textit{recipe}, based on the definitions given in the \gls{OPC UA} companion specification.
The application developer must acknowledge the match before the application can run. Then, Mendix runtime gets the results of the \gls{NN} processing, available in the address space of the respective \gls{OPC UA} server, and represents them in the dashboard, enabling real-time monitoring, as illustrated in Figure~\ref{fig:opcua_recipe}.

\section{Evaluation}
\label{section:evaluation}

This section first compares \gls{SeLoC-ML} with the traditional approach qualitatively. 
% SeLoC-ML empowers developers with \gls{TinyML}-as-a-service for cross-platform deployment. It is applicable in various scenarios and platforms, such as Arduino\footnote{\raggedright \url{https://www.arduino.cc/}} and Mentor Graphics\footnote{\raggedright \url{https://eda.sw.siemens.com/en-US/}}.
Further, a quantitative analysis is conducted. For that, the example from the last section is chosen based on our available products to quickly generate results. 

\subsection{Qualitative Analysis}

\subsubsection{Reliability and Flexibility}
Reliability is one of the essential factors in the industry since a single failure can cause significant losses. 
Unfortunately, it is not easy to achieve successful ML applications in \gls{IIoT} using the conventional workflow since many processes are involved, each of which requires extensive domain knowledge and labor. This can potentially raise the failure probability. 
% \gls{TinyML} is a highly complex subject at the intersection of data science, embedded engineering, and computer science.
% Unfortunately, it is not easy to achieve successful ML applications on the edge by the conventional workflow since many processes are involved, each of which requires extensive domain knowledge and labor.This can potentially raise the failure probability. 
\gls{SeLoC-ML} provides consistent services within one tool that can automate the engineering work by complying with semantic standards, reducing error, and guaranteeing reliability.
Moreover, it is important to provide flexible solutions to keep pace with the fast-evolving \gls{IIoT} world. \gls{SeLoC-ML} is generic enough to be applied in other scenarios, avoiding examination on a per-instance basis.

\subsubsection{Scalability and Interoperability}

It is more beneficial to reuse existing \gls{ML} models than to invent a new one every time since reusability means less cost and better scalability. However, one of the major concerns in \gls{IIoT} is heterogeneity. Numerous trained \glspl{NN} could be used for various industrial applications, but it is unclear how to apply them in a concrete use case or on specific hardware. 
As the \gls{IIoT} network expands, millions of devices from different vendors emerge, making manual management of a massive amount of hardware and software almost impossible. 
The proposed approach presents information in a unified language. This ensures that both humans and machines can consistently interpret the stored data and enable automatic development at scale. 
Besides, our semantic framework is vendor-independent and platform-neutral, enhancing the transparency and interoperability of the ecosystem.   

\subsection{Quantitative Analysis}

We quantitatively evaluate the approach using the conveyor belt example, described in Section~\ref{section:in-use-example}, since this industrial application is representative and similar results have been achieved on other use cases and platforms. 
We present the experimental results in the following steps: 1) we describe the file structure in the deployment project;
2) we compare the semantic approach (\gls{SeLoC-ML}) against the other two methods regarding the engineering effort for generating the project, scalability, error rate, and tools required. 
% Finally, we emphasize that only one tool is required for engineering an \gls{ML} application using our semantic framework, whereas other solutions involve multiple tools and demand extensive domain expertise.

% Based on the available product, we implement the use case, described in Section~\ref{section:in-use-example}, using SIMATIC S7-1500 and SIMATIC \gls{TM NPU}.
A minimum of five engineering artifacts should be engineered and created in the project: a configuration of the \gls{ML} model used, and a user logic for processing, as well as the corresponding logic in the \gls{PLC} for exchanging the data with \gls{TM NPU} and consuming its results, as shown in Table~\ref{tab:files}.

\begin{table}[ht]
      \vspace{-0.2cm}
\centering
\caption{Project files}\label{tab:files}
      \vspace{-0.2cm}
\begin{tabular}{@{}ll@{}}
\toprule
                        & Function                      \\ \midrule
\textit{npu\_app.conf}  & configure the \gls{ML} model on \gls{TM NPU} \\
\textit{main.py}        & configure on-device \gls{ML} model execution \\
\textit{DataTypes.udt}  & define data type(s) for \gls{PLC}/\gls{TM NPU} interaction \\
\textit{fbLogic.scl}    & define function block to interact with \gls{TM NPU} \\
\textit{ControlData.db} & define data block to store the data internally in \gls{PLC} memory \\ \bottomrule
\end{tabular}
      \vspace{-0.4cm}
\end{table}

We compare the implementation effort for programming our running example project for three different implementations: 1) traditional approach: programming the whole application from scratch; 2) template approach: providing a user with the pre-developed ready code template that need to be additionally configured for a specific application; 3) semantic approach: \gls{SeLoC-ML}.
% To make the comparison more intuitive, we consider the textual representation of the files, shown in Table~\ref{tab:files}. 
We count the number of \gls{LOC} needed to be manually programmed to implement the running example for every solution.
% We assume that engineer's task is to configure a running project from scratch, from an examplatory template, or using our semantic approach.
Here, we define \gls{LOC} to include the number of lines developers need to program and other configuration input that they must provide, for example, the user input in Mendix, as shown in Figure~\ref{fig:matchmaking}. 
Table~\ref{tab:loc} presents the results of measuring the engineering effort.

%produce a list of KPIs (lines of code editing, workload(clicks), steps,  error rate, …) that quantify the comparison of our workflow with the general one (Kirill’s paper can be helpful here)

% \begin{table}
% \centering
% 	\caption{Engineering effort in \gls{LOC} needed to be programmed manually based on the running example}\label{tab:loc}
% 	\begin{tabular}{|l|c|c|c|c|c|}
% 		\hline
% 		 &  npu.conf & main.py & datatypes.udt & fbLogic.scl & dataBlock.db \\
% 		\hline
% 		Traditional approach & 20 & 284 & 40 & 408 & 14 \\
% 		\hline
% 		Template application & 10 & 19 & 3 & 3 & 3 \\
% 		\hline
% 		Semantic approach & 4 & 6 & 1 & 1 & 1 \\
% 		\hline
% 	\end{tabular}
% \end{table}

\begin{table}[ht]
      \vspace{-0.2cm}
\centering
\caption{Engineering effort in \gls{LOC} based on the running example}
      \vspace{-0.2cm}
\label{tab:loc}
\begin{tabular}{@{}lcccccc@{}}
\toprule
 &
  \multicolumn{1}{l}{\textit{npu.conf}} &
  \multicolumn{1}{l}{\textit{main.py}} &
  \multicolumn{1}{l}{\textit{datatypes.udt}} &
  \multicolumn{1}{l}{\textit{fbLogic.scl}} &
  \multicolumn{1}{l}{\textit{dataBlock.db}} &
  \multicolumn{1}{l}{Total} \\ \midrule
Traditional approach   & 20 & 284 & 40 & 408 & 14 & 766 \\
Template approach & 10 & 19  & 3  & 3   & 3  & 38  \\
Semantic approach & 4  & 6   & 1  & 1   & 1  & 13  \\ \bottomrule
\end{tabular}
      \vspace{-0.4cm}
\end{table}

Moreover, we studied the flexibility of the approaches in terms of their scalability, i.e., the ability to add new data to the interaction between the \gls{PLC} and \gls{TM NPU}.
This is especially important when switching between different use cases and/or \glspl{NN}.
Compared to the traditional approach, both template and semantic approaches showed a significant reduction of the \gls{LOC}.
It is worth noting that the semantic approach scales better than the template approach, as we managed to reduce the engineering effort by a factor of three with \gls{SeLoC-ML}.

Another aspect to consider is the error rate.
Getting the most of the code generated will decrease the number of errors made during programming.
Once the code generation process is validated, the produced code will be errorless.

Additionally, we consider the number of tools needed to create the entire project for \gls{PLC} and \gls{TM NPU}. 
Using our approach, we generate the entire solution in one place using our Mendix application. 
Both traditional and template approaches require an engineer to have competencies in at least three different tools: 
a model converter tool is needed for creating the \gls{NN} configuration (\textit{npu.conf}), some \gls{IDE} for python programming to edit the user logic for \gls{NN} processing (\textit{main.py}), and \gls{TIA} Portal for \gls{PLC} programming.
Table~\ref{tab:tools} provides an overview of the tools required for each solution.

% \begin{table}
% \centering
% 	\caption{Engineering tools used for programming the running example}\label{tab:tools}
% 	\begin{tabular}{|l|c|c|c|c|c|}
% 		\hline
% 		 &  npu.conf & main .py & *.udt, *.scl, *.db \\
% 		\hline
% 		Traditional approach & Model Converter & \gls{IDE} & \gls{TIA} Portal \\
% 		\hline
% 		Template application & Model Converter & \gls{IDE} & \gls{TIA} Portal \\
% 		\hline
% 		Semantic approach & \multicolumn{3}{c|}{Mendix} \\
% 		\hline
% 	\end{tabular}
% \end{table}

\begin{table}[ht]
      \vspace{-0.2cm}
\centering
\caption{Engineering tools used for programming the running example}
      \vspace{-0.2cm}
      \label{tab:tools}
\begin{tabular}{@{}lcccc@{}}
\toprule
 & \multicolumn{1}{l}{\textit{npu.conf}} & \multicolumn{1}{l}{\textit{main.py}} & \multicolumn{1}{l}{\textit{*.udt, *.scl, *.db }} & \multicolumn{1}{l}{Total} \\ \midrule
Traditional approach   & Model Converter & Python IDE & TIA Portal & 3 \\ \cmidrule(l){2-5} 
Template approach & Model Converter & Python IDE & TIA Portal & 3 \\ \cmidrule(l){2-5} 
Semantic approach & \multicolumn{3}{c}{Mendix}                & 1 \\ \bottomrule
\end{tabular}
      \vspace{-0.5cm}
\end{table}

\section{Conclusion and Future Work}
\label{section:conclusion}

We have experienced various challenges in implementing \gls{ML} in \gls{IIoT} due to the heterogeneity of the ecosystem.
This study presents \gls{SeLoC-ML} for managing and deploying \gls{ML} on the \gls{IIoT} devices at scale by leveraging Semantic Web technology.
Many out-of-the-box features were enabled using \gls{KG}, such as knowledge discovery, similarity search, and matchmaking software (\gls{NN} models) and hardware (devices). 
By integrating \gls{SeLoC-ML} in the low-code platform, Mendix, we open new possibilities even for non-experts to easily access these semantic functionalities, use \gls{ML}-as-a-service for deploying \gls{ML} models to hardware across the platforms, and prototype end-user applications.
The ontology and code examples are available online and can be freely used and further extended. 

% Modeling heterogenous \gls{IIoT} and \gls{ML} components in a unified language is more challenging since we must balance the granularity of the semantic model with practicality. 
The next steps, which are already underway, include further improvement and integration of our approach to the production processes.
% To this purpose, we will continue to collect feedback and adapt the system from different aspects.
As illustrated in our repository, we have developed \gls{SeLoC-ML} to support other platforms than in our running example presented in the paper, such as, Arduino\footnote{\raggedright \url{https://www.arduino.cc/}}.
We intend to conduct additional analysis on other scenarios and platforms and collect feedback to further advance the robustness and scalability of our system.
% To this purpose, we will continue to collect feedback and adapt the system from different aspects.
We hope to foster the collaboration between the \gls{ML} and the Semantic Web communities.
Therefore, provisioning the framework and making the toolchain available for everyone is also one of our next steps.
% Finally, a continuous avenue of future work involves integrating \gls{IIoT} and graph \gls{NN}, facilitating the \gls{IIoT} ecosystem through knowledge-infused learning.
% \input{misc.tex}


\bibliographystyle{splncs04}
\bibliography{references}

%Author, F.: Article title. Journal \textbf{2}(5), 99--110 (2016)
%Author, F., Author, S.: Title of a proceedings paper. In: Editor,
%F., Editor, S. (eds.) CONFERENCE 2016, LNCS, vol. 9999, pp. 1--13.
%Springer, Heidelberg (2016). \doi{10.10007/1234567890}
%Author, F., Author, S., Author, T.: Book title. 2nd edn. Publisher,
%Author, A.-B.: Contribution title. In: 9th International Proceedings
%on Proceedings, pp. 1--2. Publisher, Location (2010)
%LNCS Homepage, \url{http://www.springer.com/lncs}. Last accessed 4
%\end{thebibliography}
\end{document}